\documentclass[preprint,floats,aps,showpacs,superscriptaddress]{revtex4}

\usepackage{amsmath}
\usepackage{graphicx}

\newcommand{\be}{\begin{equation}}
\newcommand{\ee}{\end{equation}}
\newcommand{\bea}{\begin{eqnarray}}
\newcommand{\eea}{\end{eqnarray}}
\newcommand{\sign}{\text{sign}}

\newcommand{\s}{\sigma}

\newcommand{\mc}{\mathcal}
\newcommand{\mcP}{\mathcal{P}}
\newcommand{\mcQ}{\mathcal{Q}}
\newcommand{\EJ}{{\rm E}_{J}}
\newcommand{\epsp}{\epsilon_+}
\newcommand{\epsm}{\epsilon_-}
\newcommand{\mrs}{m_{\rm RS}}
\newcommand{\rhorsb}{\rho_{\rm RSB}}
\newcommand{\rhof}{\rho_{\rm F}}
\newcommand{\rhofrs}{\rho_{\rm F}^{\rm RS}}
\newcommand{\rhofrsb}{\rho_{\rm F}^{\rm 1RSB}}

\begin{document}

\title{Spin glass models with ferromagnetically biased couplings on
the Bethe lattice: analytic solutions and numerical simulations}

\author{Tommaso Castellani}
\affiliation{Dipartimento di Fisica, Universit\`{a} di Roma ``La
  Sapienza'', Piazzale Aldo Moro 2, Roma I-00185, Italy}
\author{Florent Krzakala}
\affiliation{ Laboratoire P.C.T., UMR CNRS 7083, ESPCI, 10 rue
Vauquelin, 75005 Paris, France}
\author{Federico Ricci-Tersenghi}
\affiliation{Dipartimento di Fisica, Universit\`{a} di Roma ``La
  Sapienza'', Piazzale Aldo Moro 2, Roma I-00185, Italy}

\begin{abstract}
We derive the zero-temperature phase diagram of spin glass models with
a generic fraction of ferromagnetic interactions on the Bethe lattice.
We use the cavity method at the level of one-step replica symmetry
breaking (1RSB) and we find three phases: A replica-symmetric (RS)
ferromagnetic one, a magnetized spin glass one (the so-called mixed
phase), and an unmagnetized spin glass one.  We are able to give
analytic expressions for the critical point where the RS phase becomes
unstable with respect to 1RSB solutions: we also clarify the mechanism
inducing such a phase transition.  Finally we compare our analytical
results with the outcomes of a numerical algorithm especially designed
for finding ground states in an efficient way, stressing weak points
in the use of such numerical tools for discovering RSB effects.  Some
of the analytical results are given for generic connectivity.
\end{abstract}

\pacs{75.50.Lk,75.40.Cx,64.60.Fr}

\maketitle

\section{Introduction}

Spin glasses are among the most complex models in statistical
mechanics that can be treated analytically.  Even at the mean field
level their solution is highly non-trivial \cite{MEPAVI}.  Moreover
when the distribution of the disorder (the distribution of the
couplings $J_{ij}$ in the present case) is not symmetric enough, e.g.\
it has a large positive mean $\EJ[J_{ij}] > 0$, the model solution
becomes still more complex: Ferromagnetism and spin glass order may
coexist in the so-called {\em mixed phase}.

The presence of a mixed phase witnesses a very complex energy
landscape, with non-trivial thermodynamical properties.  Indeed
mean-field spin glasses are expected to have a mixed phase, while
scaling theories, like the droplet model~\cite{droplet}, do not seem
to leave any space for such a phase.

Recently the authors of Ref.~\cite{KRMA} studied numerically ground
states of the Edwards-Anderson model with an excess of ferromagnetic
couplings in 3d and they found no clear evidence for the existence of
a mixed phase.  Alternative explanations of their results, compatible
with the existence of a mixed phase, are the following: (i) the size
of the mixed phase in the 3d EA model may be very tiny; (ii) finite
size corrections may be huge and the thermodynamical limit approached
very slowly; (iii) consequences of the replica symmetry breaking may
be hard to detect in the presence of a strong bias (the global
magnetization); (iv) given a large number of quasi-ground-states (with
similar energies, but different magnetizations) the numerical
algorithm used in Ref.~\cite{KRMA} may have some small bias to find
more easily ground states with small magnetization.

In order to shed some more light on the above possible sources of
error we believe it is very useful to perform a detailed study of the
mixed phase in models where such a study can be done at an analytical
level, i.e.\ models with long-range interactions.  From the analytical
solution one can extract information on e.g.\ the size of the mixed
phase and the presence of low-energy states with different
magnetizations.  Moreover, finding ground states with the same
algorithm of Ref.~\cite{KRMA} and comparing numerical outcomes with
the analytical solution, one can study finite size effects and
possible sources of bias in the algorithm.

The complexity of the analytical solution to spin glass models with
long-range interactions depends on the interaction topology.  Those
models where each spin interacts with all the rest of the system
(fully-connected topology) can be solved in a compact way thanks to
the Parisi ansatz (see e.g.\ the recent work by Crisanti and Rizzo
\cite{CrisantiRizzoSK}).  On the contrary, the complete solution to
those models where each spin interacts only with a finite number of
neighbours (Bethe lattice topology) is much more complicated
\cite{MEPA_bethe}: Even the simplest solution with only one step of
replica symmetry breaking (1RSB) involves a functional distribution of
distributions as the order parameter.  Luckily enough such a complex
solution can be simplified in some cases: e.g.\ at zero temperature
\cite{MEPA_cavityT0}, and when sites are equivalent (factorized
solution) \cite{FLRZ} or when only zero-energy configurations are
taken into account \cite{CDMM,MERIZE}.

Spin glasses on the Bethe lattice has been extensively studied in the
second half of the eighties
\cite{VianaBray,MEPA86,Kanter,Kwon,deDomGold}.  Unfortunately at that
times it was not completely clear how to break the replica symmetry in
a way which allow for an analytical treatment: A standard replica
calculation for spin glass models on the Bethe lattice would involve
an infinite number of overlaps!  Until few years ago only replica
symmetric (RS) and variational solutions were known for spin glasses
on Bethe lattices.

The same definition of ``mixed phase'' is not clear at the RS level,
since it can not be distinguished from a non-homogeneous ferromagnet.
Indeed at the RS level only two macroscopic parameters give the full
description of the system: the magnetization $m=\sum_i m_i/N$ and the
overlap $q=\sum_i m_i^2/N$, where $m_i$ is the local magnetization on
site $i$.  Given that $q \ge m^2$, the only possible RS phases are the
following:
\begin{itemize}
\item $q=m=0$: paramagnetic phase;
\item $q>0$, $m=0$: unmagnetized spin glass phase;
\item $q=m^2>0 \Rightarrow m_i=m\;\forall i$: homogeneous ferromagnetic
phase;
\item $q>m^2>0\Rightarrow m_i$ depends on the site: Both the
non-homogeneous ferromagnetic phase and the mixed phase belong to this
class and are thus indistinguishable.
\end{itemize}
So at the RS level the presence of a mixed phase can only be deduced
from the fact that on the RS-to-RSB instability line the magnetization
is non-zero, assuming that it does not drop to zero in the RSB phase.
The only direct way of observing a mixed phase is to look for RSB
solutions with $q>m^2>0$: in this case a non-homogeneous ferromagnetic
phase corresponds to RS solutions, while a mixed phase to RSB
solutions.

In this work we will concentrate on spin glass models with a generic
fraction of ferromagnetic interactions defined on Bethe lattices with
{\em fixed} connectivity.  In order to simplify the calculations we
will perform a zero-temperature analysis of the
ferromagnetic/spin-glass transition at the level of replica symmetric
(RS) and one-step replica symmetry breaking (1RSB) solutions.

To our knowledge, the best description of the zero-temperature phase
diagram of this model is the one given by Kwon and Thouless
\cite{Kwon}.  They used a variational RS approach where the local
fields may take any real value.  However, in a model having discrete
energy levels a real-valued local field is unphysical.  Moreover,
given that our main aim is the study of the mixed phase in spin
glasses, the use of a RSB ansatz is strictly required.

Thanks to the reformulation by M\'ezard and Parisi \cite{MEPA_bethe}
of the cavity method for finite connectivity models, we are able to
derive the correct phase diagram of spin glasses with
ferromagnetically biased coupling on the Bethe lattice and to
investigate the mixed phase directly with a 1RSB ansatz.

The main questions we would like to answer are the following.  Can we
locate exactly the boundaries of the mixed phase?  How does the size
of mixed phase change with the model connectivity?  How wide or tiny
do we expect to be the mixed phase in the 3d EA model (if any)?  What
is the physical mechanism inducing the RS to RSB transition?  How
``strong'' are the measurable effects of RBS in the mixed phase?  From
the numerical point of view, how large are the finite size effects in
locating the phase transitions?  Is there any bias in the ground
states found by the numerical optimization procedure?

The work is organized as follows.  In Sec.~\ref{sec:cavity} we recall
model definition and we write cavity self-consistency equations to be
solved in Sec.~\ref{sec:k2} for Bethe lattices with fixed connectivity
3.  There we also compare numerical data to the analytic solution.  In
Sec.~\ref{sec:generic_k} we present some result valid for generic
connectivity.  Finally in Sec.~\ref{sec:discussion} the answers to
questions in the previous paragraph are discussed.

\section{The model and its solution with the cavity method at zero
temperature}
\label{sec:cavity}

We consider a 2-spin interacting spin glass model on a Bethe lattice
with fixed connectivity $c=k+1$.  The Hamiltonian of the problem is
\be
\mc{H} = -\sum_{<ij>} J_{ij} \s_i \s_j \;,
\label{H}
\ee
where $\s_i=\pm1$ are Ising variables.  The couplings $J_{ij}$ are
quenched random variables extracted from the following probability
distribution:
\be
\mathbf{P}(J) = \frac{1+\rho}{2}\;\delta(J-1) +
\frac{1-\rho}{2}\;\delta(J+1) \;.
\label{PJ}
\ee
The parameter $\rho \in[0,1]$ is thus 1 for the ferromagnet and 0 for
the unbiased spin glass.

We analyze the problem with the cavity method at zero temperature
\cite{MEPA_bethe,MEPA_cavityT0}.  The cavity method is based on the
analysis of the messages $u$ passed between sites and cavity fields
$h$ acting on each site.  Following the standard procedure one can
write self-consistency equations for the distributions of $u$s and
$h$s, that give (if the process converges in the thermodynamic limit)
the solution of the model.  The basic hypothesis of the above method is
the absence of strong correlations between two randomly chosen spins:
This is true for Bethe lattice topologies where the typical loop size
is of order $\log(N)$, which diverges in the thermodynamic limit
$N\rightarrow\infty$.

In this work we use the cavity method at two levels of approximation.
The first level corresponds to considering the system with a single
thermodynamic pure state, and it is formally equivalent to the Replica
Symmetric (RS) approach of the replica method.  The second level
corresponds to assuming the existence of many equivalent states, which
is equivalent to apply the replica method with a one step Replica
Symmetry Breaking (1RSB) approximation.

\subsection{Self-consistency equations}

For models having discrete energy levels, cavity fields at zero
temperature only take integer values, since they are related to the
difference among energy levels \cite{MEPA_cavityT0}.  Moreover for the
present Hamiltonian, given the cavity field $h$ on a site, the
corresponding message sent along the link leaving that site and having
coupling $J$ is $u=\sign(Jh)$, with the prescription that
$\sign(0)=0$.  So for any message we have that $u \in \{-1,0,1\}$.

In the RS case the self-consistency equations are
\bea
\mcP(h) &=& \int \prod_{i=1}^k d\mcQ(u_i)\;
\delta\Big(h-\sum_{i=1}^{k}u_i\Big)\;,
\label{scrs1}\\
\mcQ(u) &=& \EJ \int d\mcP(h)\;\delta\Big(u-\sign(Jh)\Big)\;,
\label{scrs2}
\eea
where $\EJ$ represent the average over the disorder distribution in
Eq.(\ref{PJ}), and $\mcP(h)$ [resp.\ $\mcQ(u)$] is the probability
distribution function (pdf) -- over the system -- of cavity fields
(resp.\ cavity messages).

The solution to RS equations can be written in terms of the
probabilities $p_0$, $p_+$ and $p_-$, defined by
\be
\mcQ(u) =  p_0\,\delta(u) + p_+\,\delta(u-1) + p_-\,\delta(u+1)\;,
\ee
with the constraint $p_0+p_++p_-=1$.

Going from the RS to the 1RSB solution \cite{MEPA_cavityT0} each field
$h_i$ (resp.\ message $u_i$) is replaced by a pdf $P_i(h)$ [resp.\
$Q_i(u)$] and the order parameters become probability distribution
functionals of pdf, $\mcP[P]$ and $\mcQ[Q]$.

The 1RSB self-consistency equations are thus
\bea
\mcP[P] &=& \int \prod_{i=1}^k D\mcQ[Q_i]\;
\delta^{(F)}\Big[P-P_0[Q_1,\ldots,Q_k]\Big]\;,
\label{scrsb1}\\
\mcQ[Q] &=& \EJ \int D\mcP[P]\;\delta^{(F)}\Big[Q-Q_0[P,J]\Big]\;,
\label{scrsb2}
\eea
where $\delta^{(F)}$ is a functional delta and the functions $P_0$ and
$Q_0$ are defined by
\bea
P_0[Q_1,\ldots,Q_k](h) &=& \frac{1}{A_k} \int \prod_{i=1}^k dQ_i(u_i) 
\; e^{-\mu\left(\sum_i|u_i|-|\sum_i u_i|\right)} \;
\delta\Big(h-\sum_{i=1}^{k}u_i\Big)\;, \label{P0} \\
Q_0[P,J](u) &=& \int dP(h)\;\delta\Big(u-\sign(Jh)\Big)\;,
\eea
where the normalization in Eq.(\ref{P0}) is given by
\be
A_k[Q_1,\ldots,Q_k] = \int \prod_{i=1}^k dQ_i(u_i) \;
 e^{-\mu\left(\sum_i|u_i|-|\sum_i u_i|\right)} \;.
\ee

The 1RSB self-consistency equations depend on the ``reweighting''
parameter $\mu$, which corresponds to the zero temperature limit of
the Parisi breaking parameter, $m \simeq \mu T$.  The solution to such
equations, as well as the corresponding zero-temperature free-energy
$\Phi$, will depend on $\mu$.

In full generality one can write the pdf of the message $u_i$ as
\be
Q_i(u_i) = \eta_0^{(i)}\,\delta(u_i) + \eta_+^{(i)}\,\delta(u_i-1) +
\eta_-^{(i)}\,\delta(u_i+1) 
\label{def_eta}
\ee
and describe the pdf $Q_i$ by the variables $\eta_0^{(i)}$ and
$\Delta\eta^{(i)}=\eta_+^{(i)}-\eta_-^{(i)}$.  So the order parameter
$\mcQ[Q]$ becomes the joint pdf $\mcQ(\eta_0,\Delta\eta)$.  Examples
of this distribution can be seen in Fig.~\ref{distr}.

\subsection{Free-energy, energy and complexity}

Since the RS solution can be formally obtained from the 1RSB one in
the $\mu \to 0$ limit, we will write only 1RSB expressions.

The free-energy $\Phi(\mu)$ is composed by two terms
\cite{MEPA_cavityT0}.  The first term, $\Phi_{\rm site}$, is computed
merging $c$ messages $u_i$, each of them having a pdf $Q_i$ randomly
extracted from $\mcQ[Q]$
\bea
\Phi_{\rm site}(\mu) &=& -\frac{1}{\mu} \int \prod_{i=1}^c
D\mcQ[Q_i]\; \log\int \prod_{i=1}^c dQ_i(u_i)\;
e^{-\mu\left(\sum_i|u_i|-|\sum_i u_i|\right)} = \nonumber\\
&=& -\frac{1}{\mu} \int \prod_{i=1}^c D\mcQ[Q_i]\; \log
A_c[Q_1,\ldots,Q_c]\; .
\eea
The second term, $\Phi_{\rm node}$, is computed in the following way
(let us write the expression for a generic $p$-spin interaction, being
$p=2$ in our case)
\bea
\Phi_{\rm node}(\mu) &=& -\frac{\EJ}{\mu} \int \prod_{i=1}^p
D\mcP[P_i]\; \log\int \prod_{i=1}^p dP_i(h_i)\;
e^{-2\mu\,\theta(-J\prod_i h_i)} = \nonumber\\ 
&=& -\frac{1}{\mu} \int D\mcP[P]\,D\mcQ[Q]\; \log\int dP(h)\,dQ(u)\;
e^{-2\mu\,\theta(-hu)}\; ,
\eea
with the prescription $\theta(0)=0$ for the step function $\theta(x)$.

The zero-temperature free-energy is given by a proper combination of
the two terms (let us write as before the expression for generic
$p$-spin interactions, being $p=2$ in our case)
\be
\Phi(\mu) = -\frac{c}{p} + \Phi_{\rm site}(\mu) - \frac{c}{p}(p-1)
\Phi_{\rm node}(\mu)\;.
\label{Phi}
\ee
Please note that Eq.(\ref{Phi}) gives the right free-energy expression
only when it is calculated with $\mcQ[Q]$ and $\mcP[P]$ solving the
self-consistency equations.

As shown in Ref.\cite{MEPA_cavityT0}, the cavity free-energy
$\Phi(\mu)$ is the Legendre transform of the complexity or
configurational entropy $\Sigma(E)$.  Then, in full analogy with
replica calculations \cite{Monasson}, one can write
\bea
E(\mu) &=& \partial_\mu \Big[ \mu\,\Phi(\mu) \Big]\; ,\\
\Sigma(\mu) &=& \mu \Big[ E(\mu) - \Phi(\mu) \Big]\; .
\eea
The complexity curve $\Sigma(E)$ can be obtained as a parametric plot
in the $\mu$ parameter.  The ground state energy of 1RSB solution is
given by the maximum of $\Phi(\mu)$, while threshold state energy is
given by the maximum of $E(\mu)$.
The solution of the present model in the spin glass phase is expected to
have an infinite number of replica symmetry breaking. In such a situation
the true threshold energy is certainly lower than its 1RSB approximation.

\section{Connectivity 3 case ($k=2$)}
\label{sec:k2}

For the ease of simplicity we will present all the details only in the
$k=2$ case, i.e.\ fixed connectivity 3, where many calculations can be
done analytically.  We leave for the next Section the results in the
generic connectivity case.

\subsection{RS solutions}

The RS equations (\ref{scrs1},\ref{scrs2}) can be written very easily
in the two variables $p_0$ and $\mrs=p_+-p_-$ (that is the
magnetization of the RS solution)
\be
\left\{
\begin{array}{l}
p_0 = p_0^2 + [(1-p_0)^2-\mrs^2]/2 \\
\mrs = \rho (1+p_0) \mrs
\end{array}
\right.
\label{eqRS}
\ee
These equations admit a paramagnetic solution with $p_0=1$ and
$p_+=p_-=0$, a spin glass solution with $p_0=p_+=p_-=1/3$, and two
ferromagnetic solutions with
\be
p_0 = \frac{1}{\rho}-1 \qquad
\mrs = \pm \sqrt{8-\frac{10}{\rho}+\frac{3}{\rho^2}} \;,
\label{ferro}
\ee
that exist only for $\rho > 3/4$.  The magnetization and the energy of
the RS solution are plotted in Fig.~\ref{mag_ener}.  From the RS
analysis one would predict solely a spin-glass/ferro transition at
$\rhofrs = 3/4$.

It was already suggested in Ref.~\cite{Kwon} that the RS phase could
be unstable for $\rho$ smaller than some $\rhorsb$.  Actually the
instability seen in Ref.~\cite{Kwon} is from integer to real-valued
fields, which is unphysical.  Nevertheless this unphysical instability
may suggests that a true instability of the RS solution towards RSB
solutions could be present.

\subsection{1RSB solutions}

In order to study the mixed phase, that is the coexistence of spin
glass order and ferromagnetic order, the replica symmetry needs to be
broken.  The mixed phase corresponds to a RSB solution, i.e.\ $0 < \mu
< \infty$, with non-zero magnetization, i.e.\ $\mcQ[Q(u)]$ not
symmetric under the transformation $u \leftrightarrow -u$.

We solve the 1RSB equations (\ref{scrsb1},\ref{scrsb2}) using a
population dynamics algorithm similar to the one used in
Ref.~\cite{MEPA_bethe}.  We evolve a population of
$(\eta_0,\Delta\eta)$ pairs, representing the joint pdf
$\mcQ(\eta_0,\Delta\eta)$, until the population becomes stationary.  A
single evolution step consists in randomly choosing $k=2$ pairs
$(\eta_0^{(1)},\Delta\eta^{(1)})$ and
$(\eta_0^{(2)},\Delta\eta^{(2)})$ from the population, and a coupling
$J$ randomly with distribution $\mathbf{P}(J)$.  Then a new pair
$(\eta_0^{(0)},\Delta\eta^{(0)})$ is generated and introduced in the
population, replacing a randomly selected pair.  The expressions for
$\eta_0^{(0)}$ and $\Delta\eta^{(0)}$ are the following
\bea
\eta_0^{(0)} &=& \frac{1}{A_2} \left[ \eta_0^{(1)}\eta_0^{(2)} +
\frac{e^{-2\mu}}{2} \left( (1-\eta_0^{(1)}) (1-\eta_0^{(2)}) -
\Delta\eta^{(1)} \Delta\eta^{(2)} \right) \right] \;, \\
\Delta\eta^{(0)} &=& \frac{J}{2 A_2} \left[ (1+\eta_0^{(1)})
\Delta\eta^{(2)} + \Delta\eta^{(1)} (1+\eta_0^{(2)}) \right] \;, \\
\text{with} \quad A_2 &=& 1 - \frac{1-e^{-2\mu}}{2}
\left( (1-\eta_0^{(1)}) (1-\eta_0^{(2)}) - \Delta\eta^{(1)}
\Delta\eta^{(2)} \right) \;,
\eea
The stationary joint pdf $\mcQ(\eta_0,\Delta\eta)$ depends on both
$\rho$ and $\mu$, giving the following scenario:
\begin{itemize}
\item for $\rho > \rhorsb \simeq 0.833$ the system is in a RS
ferromagnetic phase;
\item for $\rho < \rhorsb$ the RS solution becomes unstable towards
RSB solutions; 
\item the 1RSB solution has a non-zero magnetization as long as $\rho
> \rhofrsb \simeq 0.715$ (mixed phase) and becomes unmagnetized below
$\rhofrsb$.  In general $\rhofrsb < \rhofrs$ holds.
\end{itemize}

\begin{figure}
\begin{center}
\includegraphics[width=0.7\textwidth]{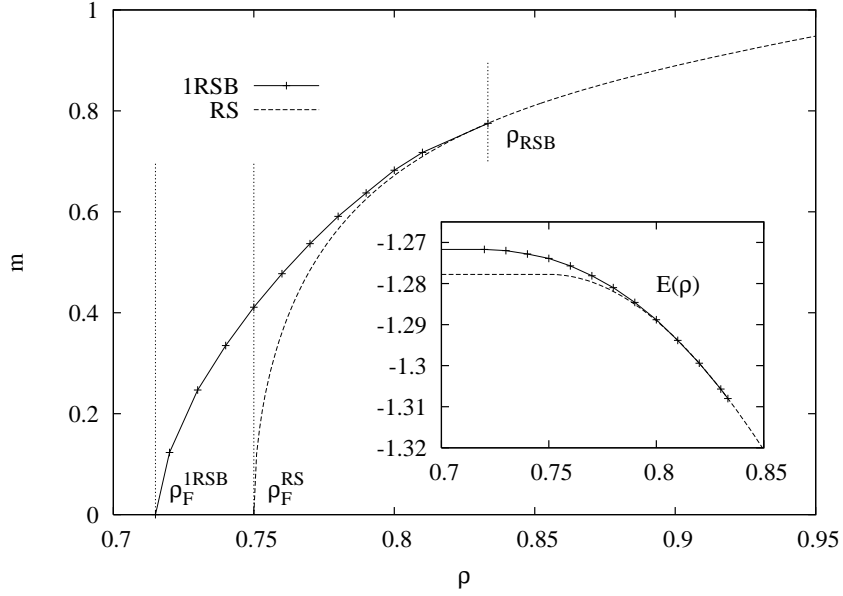}
\end{center}
\caption{Magnetization (main plot) and energy (inset) of RS and 1RSB solutions as functions of the ferromagnetic bias $\rho$.}
\label{mag_ener}
\end{figure}

In Fig.~\ref{mag_ener} we show the magnetization and the energy of
1RSB ground states, i.e.\ those states maximizing $\Phi(\mu)$: the
1RSB full curve leaves the RS dashed curve below $\rhorsb$ and becomes
$\rho$-independent below $\rhofrsb$.  We have checked that our
unmagnetized spin glass solution coincides with the one found in
previous works \cite{MEPA_cavityT0,Boettcher}.

Considering $\mu$ values different from the one maximizing $\Phi(\mu)$
we can get information also on metastable states.  We find that (i)
at the same energy level states with different magnetization do exist
and (ii) states with higher energy typically have a smaller
magnetization.

\begin{figure}
\begin{center}
\includegraphics[width=0.49\textwidth]{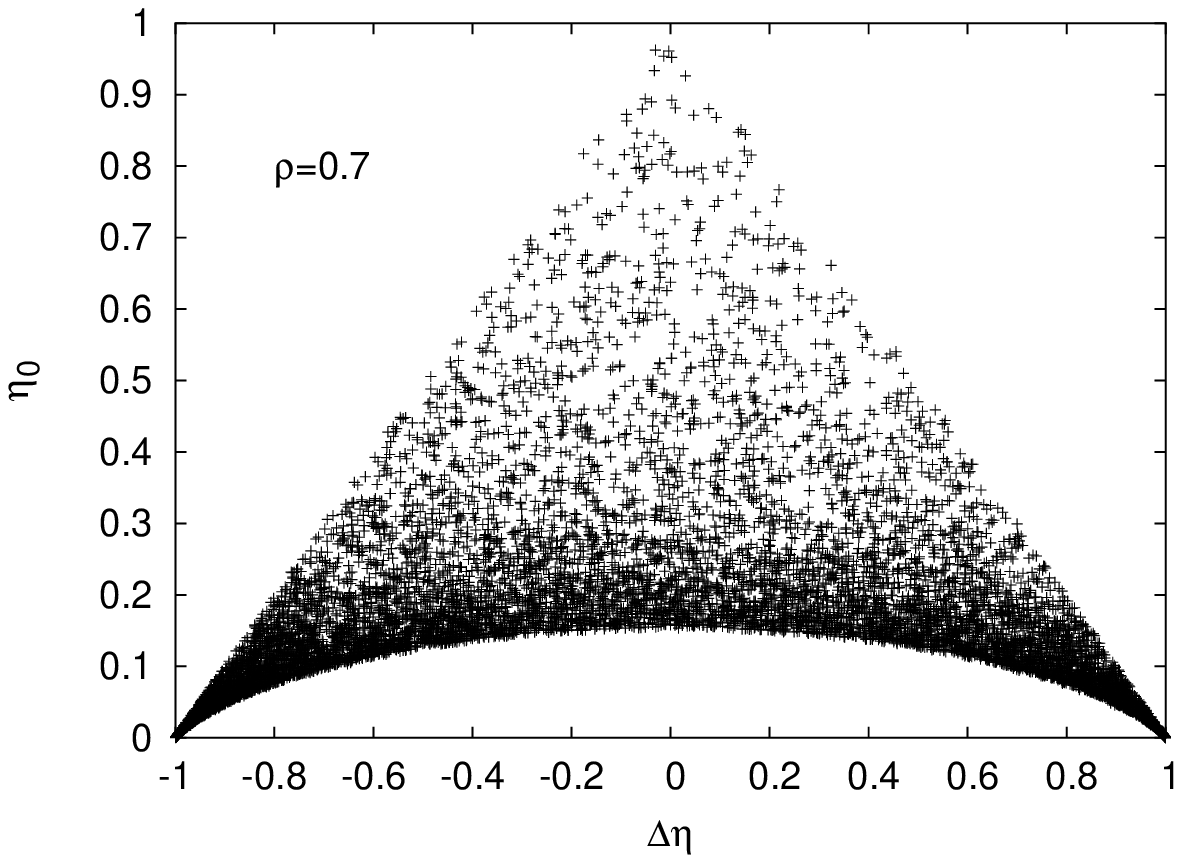}
\includegraphics[width=0.49\textwidth]{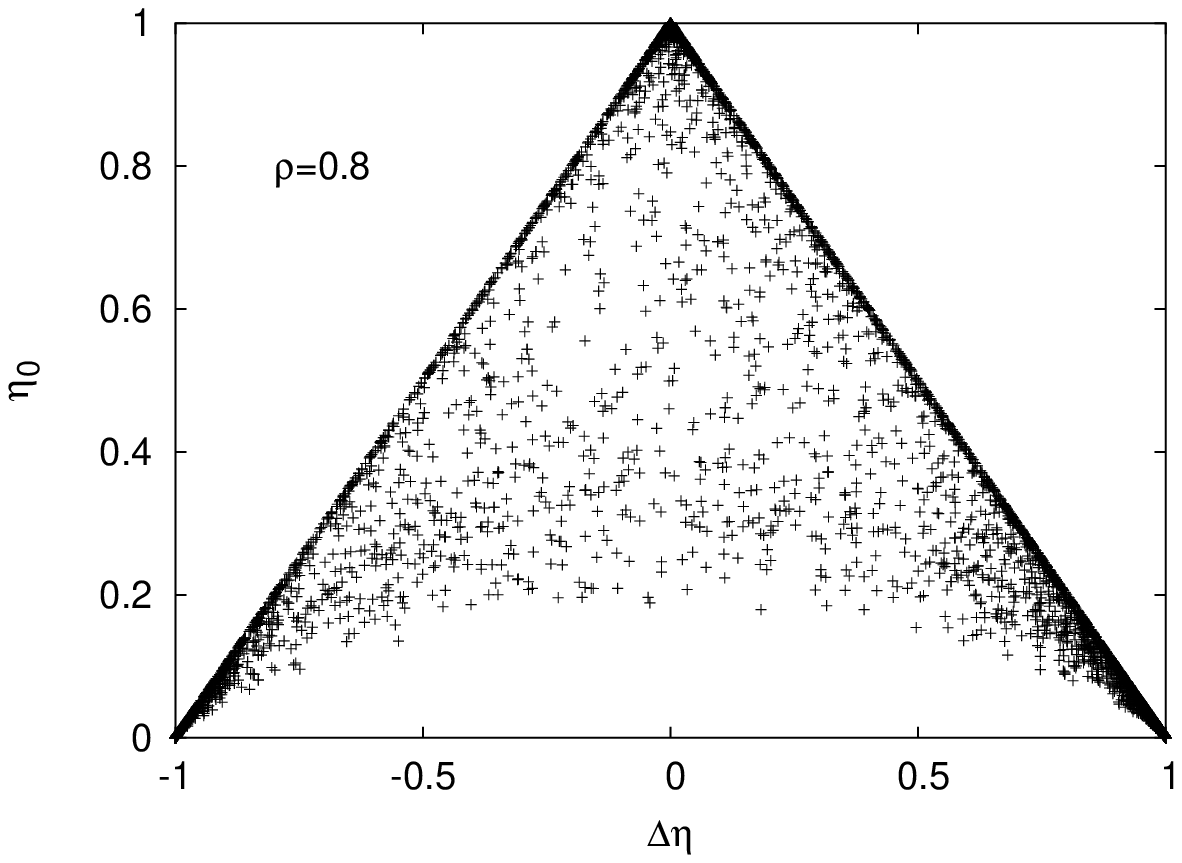}
\end{center}
\caption{Probability distributions in the unmagnetized spin-glass
(left) and the mixed phase (right).}
\label{distr}
\end{figure}

In Fig.~\ref{distr} we show the order parameter
$\mcQ(\eta_0,\Delta\eta)$ for two values of $\rho$, with $\mu$ chosen
such as to maximize $\Phi(\mu)$.  For $\rho=0.7<\rhofrsb$
(unmagnetized spin glass phase) the joint pdf
$\mcQ(\eta_0,\Delta\eta)$ is symmetric in $\Delta\eta$, that is
$\mcQ(\eta_0,-\Delta\eta) = \mcQ(\eta_0,\Delta\eta)$.  For
$\rho=0.8>\rhofrsb$ (mixed phase) the symmetry in $\Delta\eta$ breaks
down and the pdf becomes more dense around one of the bottom corners
-- the one on the right in this case.  The presence of RSB is still
clearly manifested by the spread of the points.

For $\rho > \rhorsb$ the system enters the RS ferromagnetic phase and
the order parameter $\mcQ(\eta_0,\Delta\eta)$ becomes trivial (for
this reason we do not show it), either becoming a delta function on
the point of coordinates $(p_0,\mrs)$, either concentrating on the
corners of the triangle, with weights $p_-$, $p_0$ and $p_+$ (from left
to right).

Actually we observe that approaching $\rhorsb$ from below the value of
$\mu$ maximizing the free energy diverges.  As a consequence broad
distributions are suppressed and only delta-shaped $Q(u)$ survive.
The correct RS limit is thus recovered with $\mu \to \infty$ and
$\mcQ(\eta_0,\Delta\eta)$ concentrating on the corners.

\begin{figure}
\begin{center}
\includegraphics[width=0.6\textwidth]{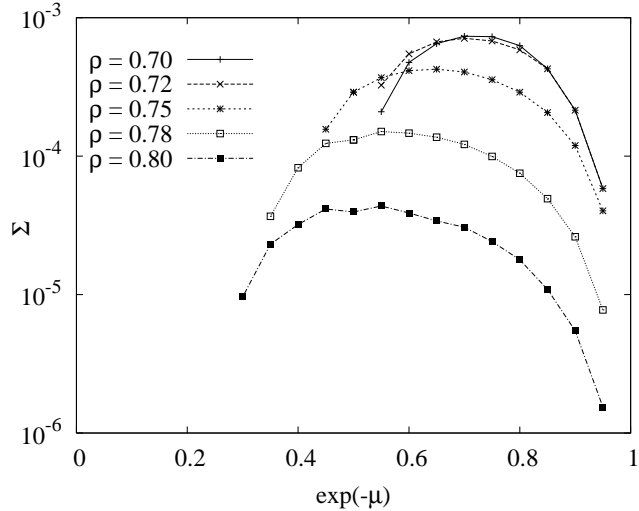}
\end{center}
\caption{The complexity $\Sigma$ as a function of $e^{-\mu}$.  For
each curve, the part corresponding to physical states is the one on
the left of the maximum.}
\label{sigma}
\end{figure}

Finally we observe in Fig.~\ref{sigma} that, increasing $\rho$ toward
$\rhorsb$, the number of states decreases rapidly (please note the
logarithmic scale): e.g.\ for $\rho=0.8$ the maximum of the complexity
$\Sigma$ is an order of magnitude smaller than for $\rho=0.72$, and
this implies that much larger system sizes have to be used in order to
detect numerically the RSB effects.

\subsection{Stability of the RS solution} 

Among the transition points that we have found with the numerical
solution of the 1RSB equations, the one in $\rhorsb$ signaling the
instability of the RS solution with respect to RSB fluctuation can be
calculated analytically.

From the 1RSB order parameter $\mcQ(\eta_0,\Delta\eta)$ one can get
back the RS solution in 2 ways
\bea
\mcQ(\eta_0,\Delta\eta) &\to& \mcQ_{\rm RS}^{(1)}(\eta_0,\Delta\eta) =
\delta(\eta_0 - p_0)\,\delta(\Delta\eta - \mrs)\;, \label{QRS1}\\
\mcQ(\eta_0,\Delta\eta) &\to& \mcQ_{\rm RS}^{(2)}(\eta_0,\Delta\eta) =
\frac{1-p_0-\mrs}{2}\,\delta(\eta_0)\,\delta(\Delta\eta+1) +
\nonumber \\
&& \qquad \qquad +\,p_0\,\delta(\eta_0-1)\,\delta(\Delta\eta) +
\frac{1-p_0+\mrs}{2}\, \delta(\eta_0)\,\delta(\Delta\eta-1) \;,
\eea
where $p_0$ and $\mrs$ satisfy the RS self-consistency equations
(\ref{eqRS}).  Our purpose is to analyze RSB fluctuations around the
above 2 solutions in close analogy to what has been done recently in
Ref.~\cite{MORI_STAB}.

Fluctuations around $\mcQ_{\rm RS}^{(1)}(\eta_0,\Delta\eta)$ are
always irrelevant.  Indeed if we replace the 2 delta function in
Eq.(\ref{QRS1}) with very narrow functions, the variances of these
functions evolve through the matrix
\be
\left(
\begin{array}{cc}
3p_0-1 & -\mrs \\
\rho\,\mrs & \rho(1+p_0) \\
\end{array}
\right)
\ee
whose eigenvalues are always less than 1 (in absolute value) as long
as $\rho > 3/4$.  This would imply a RS ferromagnetic phase always
stable.

\begin{figure}
\begin{center}
\includegraphics[width=0.28\textwidth]{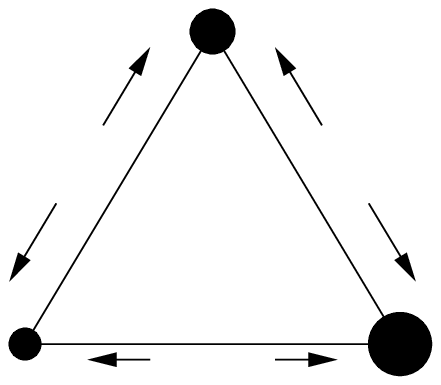}
\hspace{1cm}
\includegraphics[width=0.28\textwidth]{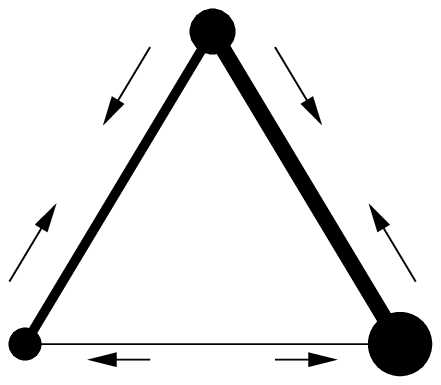}
\end{center}
\caption{Pictorial view of the 1RSB order parameter above (left) and
below (right) the instability point.  For $\rho > \rhorsb$ (left) the
distribution is concentrated on the three vertices and it is stable
under small fluctuations in any direction.  For $\rho < \rhorsb$
(right) fluctuations along the upper sides get amplified under the
population dynamics.}
\label{pictorial}
\end{figure}

However what we observe numerically is that the RS ferromagnetic state
becomes unstable around $\rhorsb \simeq 0.833$, and for $\rho$
slightly below $\rhorsb$ the population is very dense on the top sides
of the triangle (see right panel of Fig.~\ref{distr}).  This last
observation suggested us to study RSB fluctuations around $\mcQ_{\rm
RS}^{(2)}(\eta_0,\Delta\eta)$ towards a distribution of the following
kind
\begin{multline}
\mcQ(\eta_0,\Delta\eta) = \mcQ_{\rm RS}^{(2)}(\eta_0,\Delta\eta) +
\int dx\,\epsp(x)\,\delta(\eta_0-1+x)\,\delta(\Delta\eta-x)\,+ \\
+ \int dx\,\epsm(x)\,\delta(\eta_0-1+x)\,\delta(\Delta\eta+x) \;,
\label{pertQ}
\end{multline}
which is concentrated on the corners and on the top sides of the
triangle (see pictorial view in Fig.\ref{pictorial}).

In terms of pdf $Q_i(u)$,
the $\mcQ[Q]$ in Eq.(\ref{pertQ}) has the following composition
\be
Q(u)=
\left\{
\begin{array}{ll}
\delta(u) & \text{with prob. } p_0 \\
\delta(u-1) & \text{with prob. } p_+ \\ 
\delta(u+1) & \text{with prob. } p_- \\
(1-x)\,\delta(u)+x\,\delta(u-1) & \text{with prob. } \epsp(x) \\
(1-x)\,\delta(u)+x\,\delta(u+1) & \text{with prob. } \epsm(x)
\end{array}
\right.
\label{5distr}
\ee
In the following perturbative calculation, the weights on the top
sides $\epsp(x)$ and $\epsm(x)$ will be considered positive and very
small: $\epsp=\int_0^1\epsp(x)\,dx \ll 1$, $\epsm=\int_0^1\epsm(x)\,dx
\ll 1$.  In order to simplify notation let us use the following short
names for the 5 distributions in Eq.(\ref{5distr}): $\delta_0$,
$\delta_+$, $\delta_-$, $Q_+(x)$ and $Q_-(x)$.

Our purpose is to calculate how the weights $\epsp(x)$ and $\epsm(x)$
are modified under one iteration of the population dynamics, given by
the expressions in Eqs.(\ref{scrsb1},\ref{scrsb2}).  The only delicate
point is the convolution of the $k=2$ pdf $Q_1$ and $Q_2$ in
Eq.(\ref{P0}), that we write in a shorthand notation as $Q_1 * Q_2$.
A non-trivial convolution appears only when the messages $u_1$ and
$u_2$ are in contradiction, i.e.\ when they have different signs.  An
explicit calculation yields
\bea
\delta_+ * Q_-(x) &=& Q_+(f_\mu(x)) \;, \\
\delta_- * Q_+(x) &=& Q_-(f_\mu(x)) \;, \\
\text{with}\quad\qquad f_\mu(x) &=& \frac{1-x}{1-(1-e^{-2\mu})\,x} \;.
\eea
Please note that, for any finite value of $\mu$, $f_\mu$ is a
bijective map of the interval $[0,1]$ onto itself, and it has the nice
property that $f_\mu(f_\mu(x))=x$, implying that $\delta_+ *
Q_-(f_\mu(x)) = Q_+(x)$.

In a single step of the evolution dynamics, the combinations of
parents that produce a distribution $Q_+(x)$ in the population of sons
are the following: If $J=1$, $\delta_0 * Q_+(x)$ and $\delta_+ *
Q_-(f_\mu(x))$, and, if $J=-1$, $\delta_0 * Q_-(x)$ and $\delta_- *
Q_+(f_\mu(x))$.  Analogous expressions for $Q_-(x)$ can be obtained
with a $+\!\leftrightarrow\!-$ substitution.  Each one of these
expressions must be multiplied by a combinatorial factor 2.  In terms
of probabilities we have then
\bea
\epsp(x) &=& (1+\rho) \Big[\; p_0\,\epsp(x) +
p_+\,\epsm(f_\mu(x))\,|f^\prime_\mu(x)| \;\Big] + \nonumber \\
&+& (1-\rho) \Big[\; p_0\,\epsm(x) +
p_-\,\epsp(f_\mu(x))\,|f^\prime_\mu(x)| \;\Big] \;, \label{epspx} \\
\epsm(x) &=& (1+\rho) \Big[\; p_0\,\epsm(x) +
p_-\,\epsp(f_\mu(x))\,|f^\prime_\mu(x)| \;\Big] + \nonumber \\
&+& (1-\rho) \Big[\; p_0\,\epsp(x) +
p_+\,\epsm(f_\mu(x))\,|f^\prime_\mu(x)| \;\Big] \;. \label{epsmx}
\eea
The above expressions hold for continuous functions $\epsp(x)$ and
$\epsm(x)$.  In the case these function were made of delta functions,
the factors $|f^\prime_\mu(x)|$ would not appear.

In order to calculate the instability point $\rhorsb$, that is when
fluctuations along the top sides become relevant (see
Fig.~\ref{pictorial}), it is enough to consider the total weight on
the top sides $\epsp$ and $\epsm$.  Integrating
Eqs.(\ref{epspx},\ref{epsmx}) over $x \in [0,1]$, one can easily
obtain an expression for the evolution of the vector $(\epsp,\epsm)$,
given by the matrix
\be
\mc{M}=
\left(
\begin{array}{ccc}
(1+\rho)p_0+(1-\rho)p_- & & (1-\rho)p_0+(1+\rho)p_+ \\
(1-\rho)p_0+(1+\rho)p_- & & (1+\rho)p_0+(1-\rho)p_+
\end{array}
\right) \;.
\ee
Plugging into $\mc{M}$ the values of $p_0$, $p_+$ and $p_-$
corresponding to the RS ferromagnetic solution (\ref{ferro}), one
finds that the largest eigenvalue (in absolute value) of $\mc{M}$
becomes larger than 1 at $\rhorsb = 5/6 = 0.833333$.  We have thus
found an analytic expression for the instability point seen in the
numerical analysis, and the agreement is perfect.

Let us notice {\it en passant} that $\rhorsb = 5/6$ coincides with the
instability point found in Ref.~\cite{Kwon}, where the instability of
integer-valued distributions towards real-valued ones was
studied~\footnote{Please note that 2 typographic errors were present
in matrix (5.8) of Ref.~\cite{Kwon}: the first $(1-x)$ in the diagonal
terms should be replaced by $(1+x)$.}.  This integers-to-reals
instability is quite formal and in principle should not correspond to
the physical one.  However some other cases have been
found~\cite{MOPARI} where such a coincidence can be shown to exist.

Since for $\rho=\rhorsb$ the vector $(\epsp,\epsm)$ is an eigenvector
of $\mc{M}$ with eigenvalue 1, a possible set of solutions to
Eqs.(\ref{epspx},\ref{epsmx}) is given by
\be
\epsp(x)/\epsp = \epsm(x)/\epsm = S(x) = \sum_n a_n \Big[
\delta(x-x_n) + \delta(x-f_\mu(x_n)) \Big] + C(x) \;,
\ee
where $C(x)$ is a continuous function such that $C(x) =
C(f_\mu(x))\,|f^\prime_\mu(x)|,\; \forall x \in [0,1]$.

It should be then evident that the subspace of functions $S(x)$ which
become unstable under the population dynamics for $\rho<\rhorsb$
contains also functions which are very different from those
corresponding to the RS solution ($\delta_0$, $\delta_+$ and
$\delta_-$), e.g.\ the function $\delta(x-x^*_\mu)$ with $x^*_\mu =
1/(1+e^{-\mu})$ being the fixed point of the map $f_\mu$.

Moreover we observe numerically that the distribution which is
actually reached by the population dynamics is very broad, continuous
and with no delta functions.

So, in this model, the RSB instability produces an infinitesimal
fraction of distributions very different from the RS ones.  On this
aspect the instability of the present model is different from the one
studied in Ref.~\cite{MORI_STAB} for the $p$-spin model.  In that case
the RS distributions ($\delta_0$, $\delta_+$ and $\delta_-$) acquire a
small width, remaining close to the unperturbed ones.  We have also
studied this last kind of instability, and it turns out that in the
present model it becomes relevant only for $\rho<0.8265<\rhorsb$, when
the RS solution is already destabilized towards the 1RSB one.

\subsection{Comparison with numerical simulations}

In order to compare the analytical solution obtained under the 1RSB
approximation with numerically computed ground states, we have run an
algorithm analogous to the one used to obtain data in
Ref.~\cite{KRMA}. More specifically we used the Genetic
Renormalization Algorithm of~\cite{GRA}; it is a heuristic numerical
method for computing spin glass ground states with a very high level
of reliability. It uses a population based search (thus the name
genetic) and applies optimization on multiple scales (thus the name
renormalization). We have computed numerically the ground state for
Bethe lattices with fixed connectivity 3: system sizes range from
$N=64$ to $N=512$, and the number of different sample changes with $N$
in order to keep the statistical error roughly constant. For the size
studied here, errors on the ground states energy due to the heuristic
are smaller than the statistical errors~\cite{GRA}. Note however that
due to the discrete nature of the problem, there is a degeneracy of
ground state, so even if we do find a ground state with probability
almost $1$, the algorithm may have a natural bias towards some of
these ground states, as we shall discuss.

\begin{figure}
\begin{center}
\includegraphics[width=0.6\textwidth]{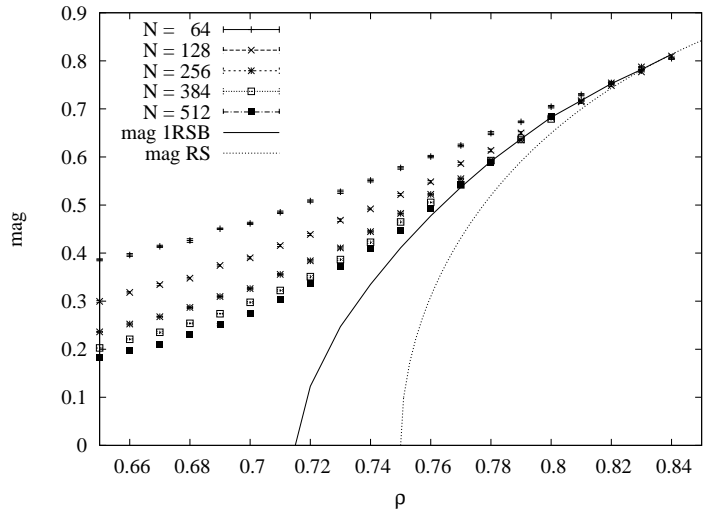}
\end{center}
\caption{Comparison between the magnetization of ground states found
  numerically (data with errors) and the analytic prediction under
  1RSB (full line) and RS (dotted line) approximations.  Statistical
  errors on numerical data is of the order of symbol size.}
\label{mag}
\end{figure}

In Fig.~\ref{mag} we compare the analytically computed magnetization
(the dotted line is the RS approximation and the full line is the 1RSB
approximation) with the magnetization of the ground states found by
the numerical algorithm.  Numerical data are far from the RS result
and are mostly compatible with the 1RSB curve.  Nevertheless, deep in
the mixed phase ($0.73 \alt \rho \alt 0.76$), we also find evidence
that the extrapolation to the thermodynamical limit of the numerically
measured magnetization is {\em below} the analytical one.  This effect
actually reduces the size of the mixed phase measured numerically.

A possible explanation to this effect is the following: given a model,
like the one we are studying here, which has many degenerate or
quasi-degenerate states with different magnetizations, an algorithm
looking for such states, starting from a trial configuration of zero
magnetization, most probably will stop in a state of magnetization
smaller than the typical one.  Such an effect has been already
observed in Ref.~\cite{Barthel}: in that case a simple algorithm for
the search of ground states always found states of zero magnetization
in a situation where the thermodynamical magnetization was non-null.

\begin{figure}
\begin{center}
\includegraphics[width=0.6\textwidth]{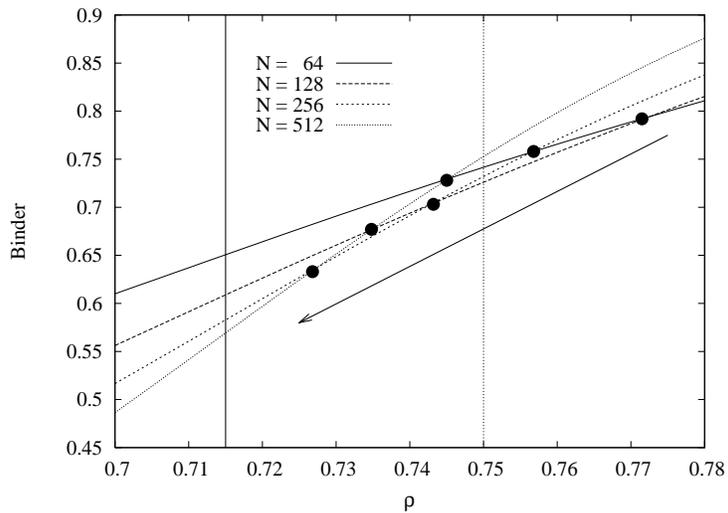}
\end{center}
\caption{Binder cumulant measured numerically for different system
 sizes do not cross at the same point, suggesting the presence of
 strong finite size effects.}
\label{binder}
\end{figure}

We also tried to deduce from the numerical data the critical point
$\rhof$ where the magnetization disappears.  In Fig.~\ref{binder} we
show the Binder parameter for sizes $N=64,128,256,512$ together with
analytical predictions (RS dotted line and 1RSB full line).
Increasing $N$, the crossing point moves to left too much in order to
be able to do any accurate prediction of $\rhof$.  This is a clear
evidence that, for this model, finite size effects are huge and make
very hard to extract information from numerical data.

Let us stress an important difference between the present model and a
similar one with Gaussian coupling, which has been studied in
Ref.~\cite{Liers} numerically and analytically at the RS level.  The
authors of Ref.~\cite{Liers} found that all the numerical results were
perfectly compatible, within the statistical error, with the analytic
predictions obtained under the RS approximation: for this reason they
concluded that RSB effect were tiny in that model.  On the contrary
here we can clearly see that numerical data are incompatible with the
RS results: e.g.\ the crossing points of Binder cumulant shown in
Fig.~\ref{binder} goes well beyond the RS critical point.

\section{Generic connectivity $k+1$}
\label{sec:generic_k}

The 1RSB equations can not be solved in a fully analytical form: even
in the simplest case ($k=2$) one needs to use a population dynamics
algorithm.  However we can compute analytically the stability point
$\rhorsb$ and the point $\rhofrs$ where the RS solution looses the
magnetization.


Let us call $S(k,r)$ the probability of having a field $h=r$ from the
sum of $k$ messages $u_1 \ldots u_k$.  Its definition is given by the
two functions $S_+(k,r)=S(k,r)$ and $S_-(k,r)=S(k,-r)$ with $r \ge 0$,
\bea
S_+(k,r) = \sum_{q=0}^{\lfloor\frac{k-r}{2}\rfloor}
\frac{k!}{q!(r+q)!(k-2q-r)!} \Big(\frac{1-p_0+\mrs}{2}\Big)^{r+q}
\Big(\frac{1-p_0-\mrs}{2}\Big)^q p_0^{k-2q-r}, \\
S_-(k,r) = \sum_{q=0}^{\lfloor\frac{k-r}{2}\rfloor}
\frac{k!}{q!(r+q)!(k-2q-r)!} \Big(\frac{1-p_0+\mrs}{2}\Big)^q
\Big(\frac{1-p_0-\mrs}{2}\Big)^{r+q} p_0^{k-2q-r},
\eea
where $\lfloor x \rfloor$ is the largest integer not greater than $x$.

For any given $k$, the self-consistency equations thus read
\bea
p_0 &=& S(k,0)\ , \label{rsk1} \\
\mrs &=& \rho \sum_{r=1}^{k} \Big[ S_+(k,r) - S_-(k,r) \Big]\ .
\label{rsk2}
\eea
Note that the right hand side of Eq.(\ref{rsk2}) is always an odd
function in the variable $\mrs$.

Equations (\ref{rsk1},\ref{rsk2}) admit a paramagnetic solution with
$\mrs=0$ and $p_0=0$, a spin glass solution with $\mrs=0$ and
$p_0=p_0^{\rm SG}$, where $p_0^{\rm SG}$ is the solution
Eq.(\ref{rsk1}) with $\mrs=0$, and a ferromagnetic solution with $\mrs
\neq 0$ and $p_0 < p_0^{\rm SG}$.

The point $\rhofrs$ where the RS magnetization vanishes can be
obtained expanding, for small $\mrs$, the right hand side of
Eq.(\ref{rsk2})
\be
\mrs = \rho\, c(p_0,k)\, \mrs + \mc{O}(\mrs^3) \ ,
\ee
and imposing the coefficient to be equal to 1 with $p_0=p_0^{\rm SG}$,
i.e.\ with the system still unmagnetized: $\rhofrs = 1 / c(p_0^{\rm
SG},k)$.

In order to compute the instability point $\rhorsb$, one can proceed
as in the previous section.  For a generic $k$ the matrix $\mc{M}$
reads
\be \label{stm}
\mc{M}=
\left(
\begin{array}{cc}
\frac{1+\rho}{2}Z + \frac{1-\rho}{2}M \quad &
\frac{1-\rho}{2}Z + \frac{1+\rho}{2}P \\
\frac{1-\rho}{2}Z + \frac{1+\rho}{2}M \quad &
\frac{1+\rho}{2}Z + \frac{1-\rho}{2}P
\end{array}
\right)
\ee
with $Z=k\,S(k-1,0)$, $P=k\,S_+(k-1,1)$ and $M=k\,S_-(k-1,1)$.  The
instability point $\rhorsb$ corresponds to the value of $\rho$ for
which the maximum eigenvalue of $\mc{M}$ becomes larger than 1 in
absolute value.

Let us eventually observe that the function $S(k,0)$ is a polynomial
in $p_0$ with only powers of the same parity of $k$.  This implies
that, for odd $k$, the solution $p_0=0$ always exist, but it is stable
under a small perturbation in $p_0$ only for $\rho < \rho_0$.  The
stability point $\rho_0$ can be easily computed as the value where the
coefficient of the linear term in $S(k,0)$ is equal to 1.  Analogously
to what has been found in $p$-spin models \cite{MORI_STAB}, we find
that for odd $k$ the equality $\rhorsb = \rho_0$ always holds.

\begin{figure}
\begin{center}
\includegraphics[width=.49\textwidth]{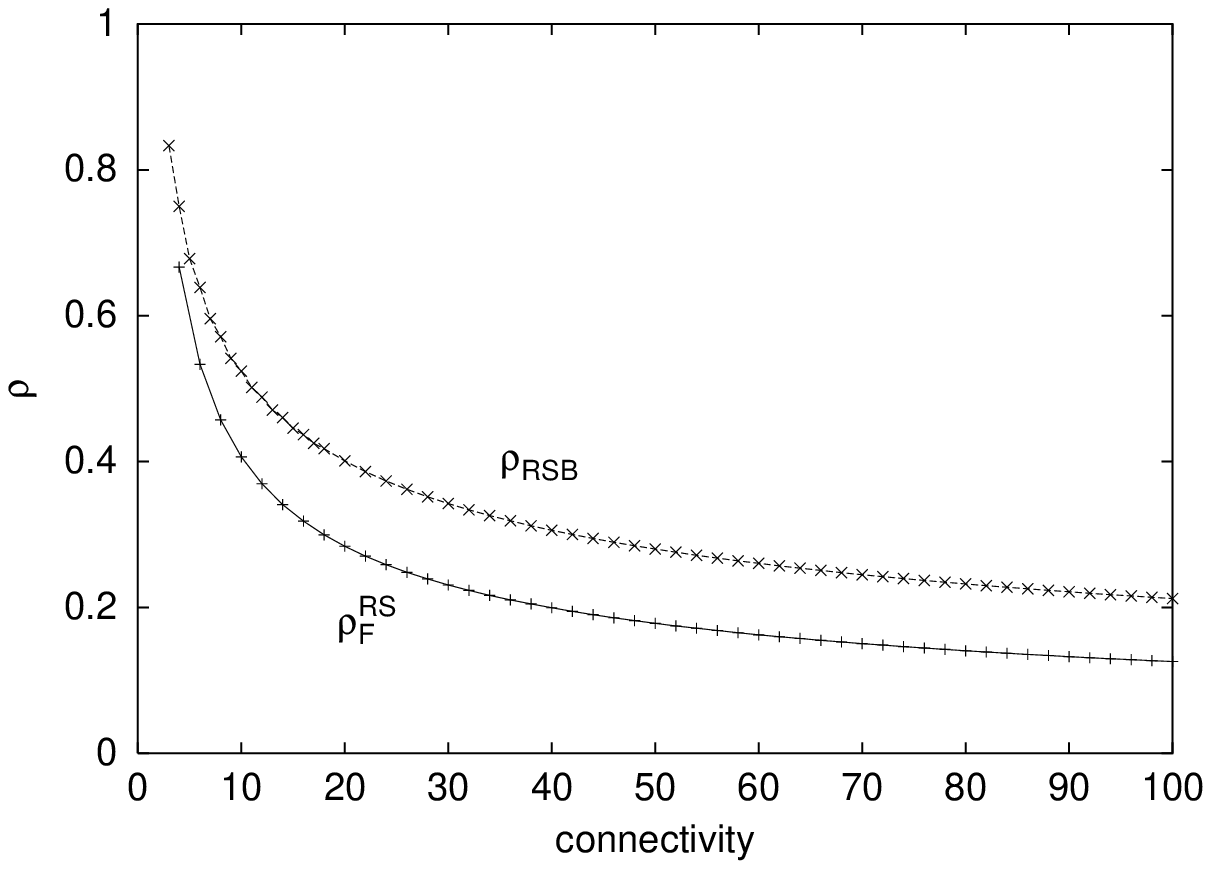}
\includegraphics[width=.49\textwidth]{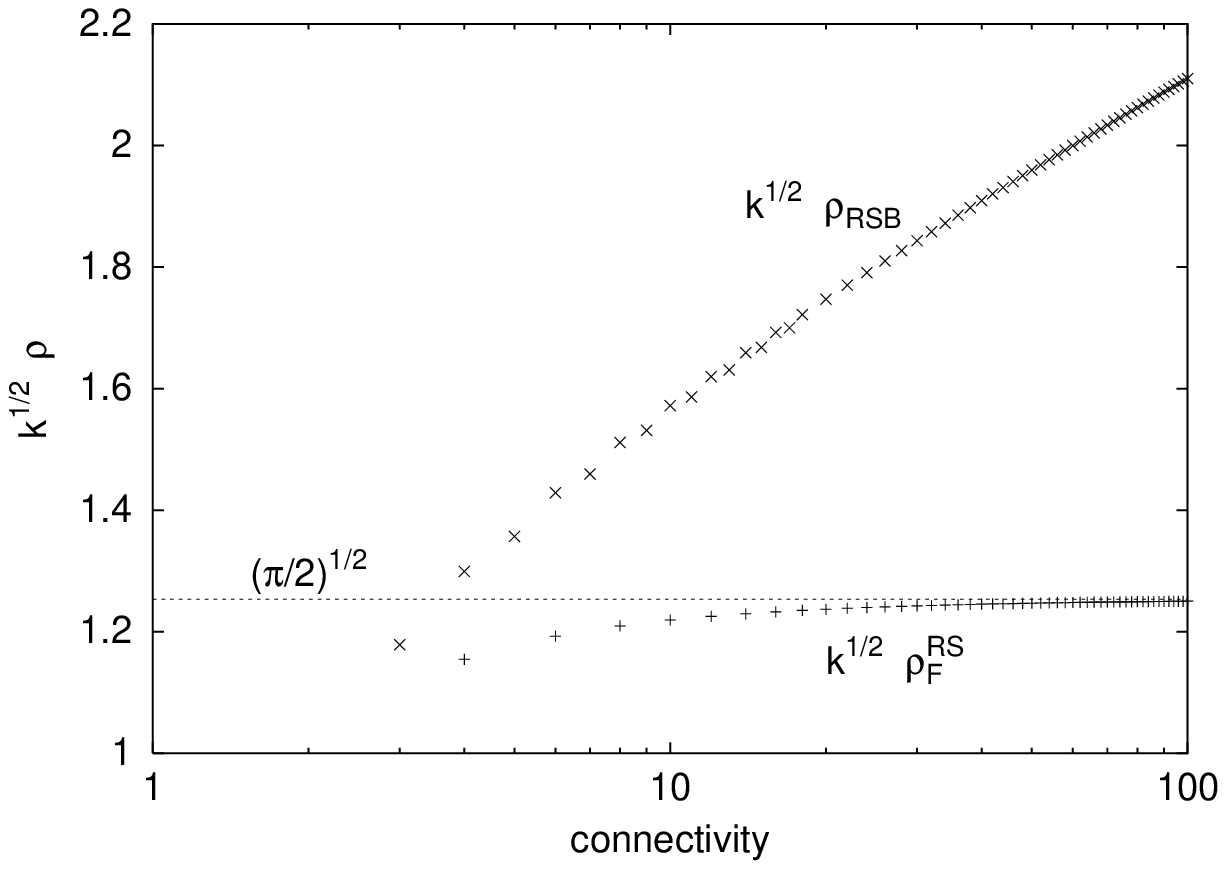}
\end{center}
\caption{Left plot: The RSB instability point $\rhorsb$ and the point
$\rhofrs$ where the RS solution becomes unmagnetized, as a function of
the connectivity $k\!+\!1$.  Right plot: rescaled variables.}
\label{rhok}
\end{figure}

In Fig.~\ref{rhok} we summarize the values of $\rhorsb$ and $\rhofrs$
for many values of the connectivity $k+1$.  It is easy to check that
$\rhorsb > \rhofrs$ strictly for any connectivity, and that $\rhorsb
\sim \log(k)/\sqrt{k}$ and $\rhofrs \sim 1/\sqrt{k}$ for $k \gg 1$.

For all the values of $\rho$ that we checked, the magnetization of a
solution is a non-monotonous function of the reweighting parameter
$\mu$.  When plotted versus $\exp(-\mu)$ it has a parabolic shape,
taking its maximum very close to the value of $\mu$ maximizing
$\Phi(\mu)$ and its minima at the extrema $\mu=0$ and $\mu=\infty$
corresponding to the RS solution.  So the magnetization in the RSB
solutions is typically larger than in the RS solution (see also
Fig.~\ref{mag_ener} for the $k=2$ case).  This implies the inequality
$\rhofrsb \le \rhofrs$, and the size of the mixed phase is bounded
from below by the quantity $(\rhorsb-\rhofrs)$, which is strictly
positive for any finite connectivity.

We have also checked that in the $k\to\infty$ limit our results
converge to those for the SK model \cite{SK}.  Indeed for odd $k$ and
$\rho < \rhorsb=\rho_0$ the solution has $p_0=0$, and the expression
for $\rhofrs$ simplifies to
\begin{equation}
\rhofrs = \left[ \sum_{i=0}^{(k-1)/2} \frac{k!}{i!(k-i)!}\;
\frac{k-2i}{2^{k-1}} \right]^{-1} \quad .
\end{equation}
In the $k \to \infty$ limit, $\rhofrs \sim \sqrt{\pi/(2k)}$ which
corresponds to the value found by De Almeida and Thouless~\cite{AT},
once the energy of the system is rescaled by the proper $\sqrt{k}$
factor.

\begin{figure}
\begin{center}
\includegraphics[width=0.6\textwidth]{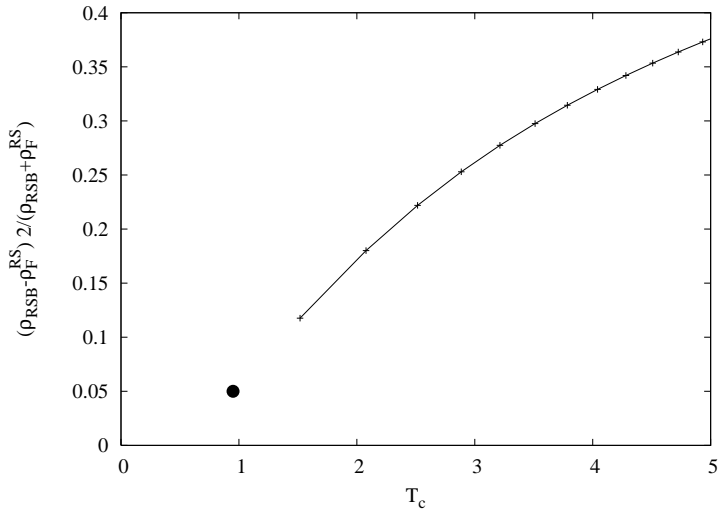}
\end{center}
\caption{The relative size of the mixed phase for a spin glass on the
  Bethe lattice as a function of the critical temperature. The big
  black dot is the result for the 3d EA model (after \cite{KRMA}).}
\label{rel_size}
\end{figure}

The authors of Ref.~\cite{KRMA} reported that, if any mixed phase
existed in the 3d EA model, its size would be very tiny: defining the
relative size as
\begin{equation}
\frac{\rhorsb - \rhof}{(\rhorsb+\rhof)/2}\;,
\end{equation}
their numerical findings are compatible with a relative size of 0.05
roughly.  In order to compare such a numerical result for the 3d EA
model with the analytical estimation of the mixed phase found here, we
plot in Fig.~\ref{rel_size} the relative size of the mixed phase for
different connectivities as a function of the critical temperature.
The black big point in Fig.~\ref{rel_size} corresponds to the
numerical result for the 3d EA model.  We see clearly that such a
numerical finding is perfectly compatible with the mean-field
prediction. Again, some 4d simulations are necessary to confirm or
infirm the absence of a mixed phase in finite dimensions.

\section{Summary and discussion}
\label{sec:discussion}

In this work we have studied analytically and numerically the low
temperature phase of a spin glass model with ferromagnetically biased
couplings defined on a Bethe lattice with fixed connectivity.  We have
shown that such a model has a mixed phase for any connectivity and
that the relative size of such a mixed phase may change a lot with the
connectivity.  Exact locations of the phase boundaries have been
computed numerically and even analytically when possible (e.g.\
$\rhorsb$ for even connectivity).  The instability which induces the
spontaneous breaking of the replica symmetry has been deeply analyzed.

Regarding the lack of a clear evidence for a mixed phase in the 3d EA
model we have found many reasons for that.  Under the Bethe
approximation, the expected size of the mixed phase in 3d is very tiny
and perfectly compatible to what has been found numerically in
Ref.~\cite{KRMA} (see Fig.~\ref{rel_size}).  Finite size corrections
on the numerical data are huge (see Fig.~\ref{binder}).  Consequences
of the RSB may be very hard to detect numerically with small systems:
e.g.\ the complexity in the mixed phase may be very small (see
Fig.~\ref{sigma}).  Finally the same algorithm may have some small
bias, whose main effect is to reduce the size of the mixed phase.  For
all these reasons we believe that the study of the mixed phase is in
general, as the study of the spin glass phase in field~\cite{AT}, a
very difficult task from the numerical point of view.

Let us finish discussing a point that we believe interesting: the
physical meaning of the equality $\rhorsb = \rho_0$.  For even
connectivities the number of cavity messages arriving on a site is an
odd number.  So the solution with no null messages always exists.  It
is very curious that the RS-to-RSB instability studied here (as well
as the 1RSB-to-2RSB instability studied in Ref.~\cite{MORI_STAB})
coincides with the appearance of null messages.  This coincidence
strongly suggests that (for even connectivities) null messages are
unphysical and they arise only when the solution ceases to be the
correct one.  The instability of the 1RSB solution toward further
steps of replica symmetry needs to be studied in order to check the
above conjecture.

\begin{acknowledgments}
We acknowledge the financial support provided through the European
Community's Human Potential Programme under contracts
HPRN-CT-2002-00319, Stipco and HPRN-CT-2002-00307, Dyglagemem.
\end{acknowledgments}


\begin{thebibliography}{99}

\bibitem{MEPAVI} M.~M\'{e}zard, G.~Parisi, M.A.~Virasoro, \textit{Spin
Glass Theory and Beyond}, World Scientific, Singapore (1987)

\bibitem{droplet}  A. Bray and M. Moore, J. Phys. C {\bf 17}, L463
  (1984). D. Fisher and D. Huse, Phys. Rev. B {\bf 38}, 373 (1988).

\bibitem{KRMA} F.~Krzakala and O.C.~Martin, Phys.~Rev.~Lett.
\textbf{89}, 267202 (2002).

\bibitem{CrisantiRizzoSK} A.~Crisanti, T.~Rizzo, Phys.~Rev.~E {\bf
65}, 046137 (2002).

\bibitem{MEPA_bethe} M.~M\'{e}zard, G.~Parisi, Eur.~Phys.~J.~B {\bf
20}, 217 (2001).

\bibitem{MEPA_cavityT0} M.~M\'{e}zard, G.~Parisi, J.~Stat.~Phys. {\bf
111} 1 (2003).

\bibitem{FLRZ} S.~Franz, M.~Leone, F.~Ricci-Tersenghi, and
R.~Zecchina, Phys.~Rev.~Lett. \textbf{87}, 127209 (2001).

\bibitem{CDMM} S.~Cocco, O.~Dubois, J.~Mandler, and R.~Monasson,
Phys.~Rev.~Lett. \textbf{90}, 047205 (2003).

\bibitem{MERIZE} M.~M\'ezard, F.~Ricci-Tersenghi, and R.~Zecchina,
J.~Stat.~Phys. \textbf{111}, 505 (2003).

\bibitem{VianaBray} L.~Viana and A.J.~Bray, J.~Phys.~C \textbf{18},
3037 (1985).

\bibitem{MEPA86} M.~M\'ezard and G.~Parisi, Europhys.~Lett.
\textbf{3}, 67 (1987).

\bibitem{Kanter} I.~Kanter and H.~Sompolinsky, Phys.~Rev.~Lett.
\textbf{58}, 164 (1987).

\bibitem{Kwon} C.~Kwon and D.J.~Thouless, Phys.~Rev.~B {\bf 37}, 7649
(1988).

\bibitem{deDomGold} C.~de~Dominicis and Y.Y.~Goldschmidt, J.~Phys.~A
\textbf{22}, L775 (1989).

\bibitem{Monasson} R.~Monasson, Phys.~Rev.~Lett. {\bf 75}, 2847
(1995).

\bibitem{Boettcher} S.~Boettcher, Eur.~Phys.~J.~B \textbf{31}, 29
(2003).

\bibitem{MORI_STAB} A.~Montanari and F.~Ricci-Tersenghi,
Eur.~Phys.~J.~B \textbf{33}, 339 (2003).

\bibitem{MOPARI} A.~Montanari, G.~Parisi, and F.~Ricci-Tersenghi,
J.~Phys.~A \textbf{37}, 2073 (2004).

\bibitem{GRA} J. Houdayer, O. C. Martin, Phys. Rev. Lett. 83 (1999)
  1030-1033 and Phys. Rev. E {\bf 64}, 056704 (2001).

\bibitem{Barthel} W.~Barthel, A.K.~Hartmann, M.~Leone,
F.~Ricci-Tersenghi, M.~Weigt, and R.~Zecchina, Phys. Rev. Lett. {\bf
88}, 188701 (2002).

\bibitem{Liers} F.~Liers, M.~Palassini, A.~K.~Hartmann and M.~Juenger,
Phys.~Rev.~B \textbf{68}, 094406 (2003).

\bibitem{SK} D.~Sherrington and S.~Kirkpatrick, Phys. Rev. Lett. {\bf
32} 1792 (1975).

\bibitem{AT} J.R.L.~de~Almeida and D.J.~Thouless, J. Phys. A {\bf 11},
983 (1977).

\end{thebibliography}
\end{document}